\documentclass[preprint]{aastex}  

\usepackage{amsmath}
\usepackage{amssymb}
\usepackage{graphicx}
\usepackage{natbib}

\renewcommand{\b}[1]{\boldsymbol{#1}} 
\newcommand{\unit}[1]{\nobreak{\mathrm{\;#1}}} 
\newcommand{\mr}[1]{\nobreak{\mathrm{#1}}} 
\newcommand{\msc}[1]{\nobreak{\textsc{#1}}} 
\newcommand{\ud}{{\rm d}} 

\newcommand{\bA}{\boldsymbol{A}}
\newcommand{\vv}{\boldsymbol{v}}
\newcommand{\bH}{\boldsymbol{H}}
\newcommand{\bomega}{\boldsymbol{\omega}}
\newcommand{\be}{\boldsymbol{e}}
\newcommand{\br}{\boldsymbol{r}}
\newcommand{\bn}{\boldsymbol{n}}
\newcommand{\grad}{\boldsymbol{\nabla}}
\newcommand{\cross}{\boldsymbol{\times}}

\begin{document}

\title{Production of magnetic energy by macroscopic turbulence in GRB afterglows}
\author{Lorenzo Sironi\\ Jeremy Goodman}
\affil{Princeton University Observatory, Princeton, NJ 08544}
\email{jeremy@astro.princeton.edu}

\begin{abstract}
  Afterglows of gamma-ray bursts are believed to require magnetic fields
much stronger than that of the compressed pre-shock medium.
As an alternative to microscopic plasma instabilities,
we propose amplification of the field by macroscopic turbulence
excited by the interaction of the shock with a clumpy pre-shock medium, for
example a stellar wind.  Using a recently developed formalism for localized
perturbations to an ultra-relativistic shock,
we derive constraints on the lengthscale, amplitude,
and volume filling factor of density clumps required to produce a given
magnetic energy fraction within the expansion time of the shock, assuming
that the energy in the field achieves equipartion with the turbulence.  Stronger
and smaller-scale inhomogeneities are required for larger shock Lorentz factors.
Hence it is likely that the magnetic energy fraction evolves as the shock slows.
This could be detected by monitoring the synchrotron cooling frequency if the
radial density profile ahead of the shock, smoothed over clumps, is known.

\end{abstract}

\maketitle

\section{Introduction}

Since their discovery, gamma-ray burst afterglows have been attributed
to synchrotron radiation from the forward shock wave
\citep{meszaros_rees97}, although it has been recently argued
\citep{uhm_beloborodov07, genet_etal07} that observations might
support a model in which the forward shock is invisible and the
afterglow is emitted by a long-lived reverse shock in the burst
ejecta. Assuming anyway a forward-shock origin for the afterglow emission, it
is difficult to account for the magnetic energy density behind
the forward shock by simple compression of the pre-shock field.
Interstellar magnetic energy densities are typically comparable to
thermal pressures and are therefore a fraction
$\epsilon_{\msc{b},0}\equiv \rho_{\mr{mag},0}/\rho_0=10^{-9}-10^{-7}$ of the
total internal energy density when rest mass is included.  It is
possible that the pre-shock medium is the stellar wind of the burst
progenitor; while the magnetic energy fraction in winds is less well
known, it is unlikely to be much larger than this.  Simply compressing
the medium would produce approximately the same ratio
$\rho_{\mr{mag}}/\rho$ behind the shock.  Instead, phenomenological
models of afterglow light curves typically require
$\epsilon_{\msc{b}}= 10^{-3}-10^{-1}$ \citep{panaitescu_kumar02,
  yost_etal03, panaitescu05}.  It follows that the magnetic energy per
baryon must be increased by $\sim 10^4-10^8$.

In fact, gamma-ray burst (hereafter GRB) afterglows present the most
compelling case among astrophysical collisionless shocks for prompt
creation of magnetic energy.  The synchrotron emission from supernova
remnants is generally consistent with compression of the interstellar
field, although some modest additional amplification may be required
in particular cases \citep{volk_etal05}.  We focus here on GRB
afterglows rather than GRB internal shocks, which also emit by some
combination of synchrotron and synchrotron self-Compton, because it
may be that the burst ejecta are magnetically dominated
\citep{coburn_boggs03, zhang_etal03}, and so the internal shock emission is driven by the progenitor magnetic field.

The leading hypothesis for field amplification in GRB afterglows is
the relativistic Weibel instability, which extracts free energy from
the anisotropy of the particles' velocity distribution function,
producing filamentary currents aligned with the shock normal; these
currents are responsible for the creation of transverse magnetic
fields \citep{medvedev_loeb99}.  This process is able to violate MHD
flux-freezing because it occurs on a microscopic scale---the
relativistic electron or ion skin depth---where the inertia of
individual charged particles is significant.  While the Weibel
instability provides a plausible mechanism to isotropize the particle
velocities, it is unclear whether the small-scale fields that it
produces can survive mutual annihilation long enough to explain the
observed synchrotron afterglow emission.  Several groups
\citep{silva_etal03, frederiksen_etal04, spitkovsky_05} have attempted
to simulate the long-term nonlinear outcome of the instability, but a
consensus on this question has not been achieved yet \citep[and
references therein]{waxman06}.  One might have thought that if this
instability were the source of post-shock fields, then
$\epsilon_{\msc{b}}$ should have a universal value for highly
relativistic, highly collisionless shocks.  Yet, while
$\epsilon_{\msc{b}}$ is modeled by a constant for individual GRB
afterglows, it seems to vary from one afterglow to another
\citep{panaitescu_kumar02, yost_etal03, panaitescu05}.

In this paper, we explore a traditional magnetohydrodynamic
explanation for magnetic field growth: turbulence.  It is well known
in non-relativistic fluid dynamics that oblique shocks produce or
alter the vorticity of a fluid \citep{ishizuka64}. In this paper we
will show that the same is true for an ultra-relativistic shock
passing over density inhomogeneities in the pre-shock circumburst
medium. The formalism described in a previous paper
\citep{goodman_macfadyen07} has let us define the vorticity created in
an ultra-relativistic fluid in which the energy-momentum tensor can be
approximated by that of an ideal fluid with pressure equal to one
third of the proper energy density ($P=\rho/3$). In the same work, we
have introduced a remarkably simple but accurate general approximation
for the local modulation of the shock Lorentz factor ($\Gamma$) by
pre-shock density inhomogeneities; within this approximation, it is
not necessary to follow the details of the flow far downstream in
order to predict the evolution of the shock, provided that $\Gamma\gg
1$, that the pre-shock pressure is negligible and that the post-shock
pressure satisfies $P=\rho/3$.  This approximation, which is modeled
on non-relativistic results described by \citet{whitham74}, reproduces
exactly the self-similar evolution of $\Gamma$ for a shock advancing
into a cold pre-shock medium with a power-law density profile in
planar, cylindrical, or spherical symmetry \citep{Sari06a}.  More
importantly for the present purpose, it allows us to estimate the
post-shock vorticity resulting from a prescribed pre-shock density
that varies along as well as perpendicular to the shock normal.  Given
the vorticity, we divide the post-shock velocities, which are
marginally non-relativistic in the average post-shock rest frame, into
vortical and non-vortical parts.  We presume that the energy density
in vortical motions is a measure of the magnetic energy density that
will eventually result after the eddies wind up the field to the point
where its backreaction on the turbulence becomes important.  These
methods are described in \S II.

In \S III, we briefly review the present state of knowledge concerning
the density inhomogeneities that may exist ahead of the shock.  Both
the amplitude and the lengthscale of the inhomogeneities are
important.  The former controls the amount of vortical energy---and
then of magnetic energy---that is produced, while the latter
determines the eddy-turnover time of the turbulence, which---when
multiplied by the number of eddy rotations necessary to amplify the
field up to the observed $\epsilon_{\msc{b}}$---must be less than the
shock deceleration time, so that the field can be significantly
amplified before adiabatic expansion reduces the particle energies
available to be radiated.  The uncertainties are large because one
doesn't know whether the pre-shock medium is more like a stellar wind
or like some component of the Galactic interstellar medium, and
because the lengthscales of interest are too small ($\lesssim 10^{14}
\unit{cm}$) to be directly resolved even in the interstellar medium.
Inhomogeneities on somewhat larger scales have been invoked to explain
undulations in afterglow light curves \citep{Wang_Loeb00,
  lazzati_etal02, schaefer_etal03, Nakar_Piran_Granot_03}.

In \S IV, we use the formalism of \S II to characterize the density
contrasts and lengthscales that pre-shock clumps should have in
order to amplify the magnetic field up to the observed value, in the
light of the circumburst picture outlined in \S III. We find that, for
smaller shock Lorentz factors, the constraints on clump sizes and
overdensities become less stringent; as a consequence, the magnetic
energy fraction produced by pre-shock clumps \emph{via} macroscopic
turbulence is expected to evolve as the shock slows down. In \S V, we
comment on the plausibility of our proposed mechanism to explain the
magnetic field amplification in GRB afterglows and we discuss how the
results obtained in \S IV could be tested by inferring the time
dependence of $\epsilon_{\msc{b}}$ from the time evolution of the
observed synchrotron cooling frequency as the shock ages.

\section{Geometrical Shock Dynamics}\label{sec:CCW}

The evolution of a shock advancing into an inhomogeneous medium
depends, in principle, upon the details of the downstream flow behind
the shock and of the ``piston'' that drives it.  Geometrical Shock
Dynamics (hereafter GSD) is an approximation for this evolution in
which only the conditions at the shock appear explicitly.  Originally
formulated by \citet{whitham74} for non-relativistic fluids, GSD has
been extended by \citet[hereafter Paper I]{goodman_macfadyen07} to
strong ($\Gamma\gg1$) ultra-relativistic shocks advancing into an
ideal fluid whose pressure is negligible ahead of the shock, but one
third of its proper energy density behind the shock ($P=\rho/3$).

The fundamental approximation of GSD is to evaluate the forward-going
Riemann characteristics in the post-shock flow as if that flow were (i)
isentropic and (ii) homogeneous far downstream with properties
determined by the mean shock speed and the mean pre-shock density.
Actually, pre-shock density inhomogeneities lead to post-shock entropy
variations, so assumption (i) is wrong in principle, but it turns out
to be a useful fiction.  In this respect, an ultra-relativistic flow has
the advantage that since pressure depends only on energy density and
not on any other thermodynamic variable (such as the proper number
density of baryons, $N$), the actual entropy is irrelevant to the
Riemann characteristics, which therefore enjoy exact Riemann
invariants.  So assumption (i) is well justified except insofar as it
may be compromised by secondary shocks created by the inhomogeneities
themselves behind the main shock.  Assumption (ii) is
reasonable when pre-shock density inhomogeneities are small in
lengthscale, so that they may be expected to average out far
downstream.

With assumption (i), the Riemann invariant on the forward
characteristics has the same value just behind the shock as it does
far downstream, and therefore, with assumption (ii), the same value
that it would have in the mean flow.  Together with the jump
conditions across the shock, this provides a relation between the
local Lorentz factor of the shock, $\Gamma$, and the proper pre-shock
energy density $\rho_0\approx mN_0 c^2$, in which $N_0$ is the proper
number density of nucleons ahead of the shock and $m$ is the rest mass
per nucleon.  In the essentially one-dimensional case that $\rho_0$
varies along the shock normal but not perpendicular to it, the
ultra-relativistic GSD relation for the response of the shock Lorentz
factor $\Gamma$ to localized and transitory variations in the
pre-shock density $\rho_0$ becomes (Paper I)
\begin{equation}\label{eq:gamma}
\Gamma=\bar\Gamma\left(\frac{\rho_0}{\bar\rho_0}\right)^{-\lambda}\qquad
\mbox{where}~\lambda\equiv\sqrt{3}-\frac{3}{2}\approx0.232 
\end{equation}
and the corresponding change in the post-shock pressure is 
\begin{equation}\label{eq:p}
P=\bar P\left(\frac{\rho_0}{\bar\rho_0}\right)^{1-2\lambda}\approx
\bar{P}\left(\frac{\rho_0}{\bar\rho_0}\right)^{0.536}\,,
\end{equation}
where the overbars indicate mean values.  As in non-relativistic GSD,
these relations can be extended to multidimensional flows in which
$\rho_0$ varies laterally as well as longitudinally (with respect to
the shock normal), causing convergence or divergence of the shock
normals.  Relations (\ref{eq:gamma}) and (\ref{eq:p}) are then
modified by factors involving the ratio of the local shock area to its
mean value (Paper I).  It is shown, however, that these corrections
are of higher order in $\Gamma^{-1}$ unless the density contrasts are
$\sim O(\Gamma)$.  For the conditions contemplated in this paper,
eqs.~(\ref{eq:gamma}) and (\ref{eq:p}) will be adequate even in two or
three dimensions.

As in the original non-relativistic theory, rigorous error estimates
for ultra-relativistic GSD are difficult.  Informally, the following
conditions are probably necessary for the approximation to be useful.
First, the pre-shock medium should be cold, meaning that pre-shock
pressure satisfies $P_0\ll\rho_0$ and that internal velocities are
$\ll c$; this is very likely true of the external forward shocks of
GRBs.  Second, the lengthscales of the pre-shock inhomogeneities
should be small compared to the shock radius, so that the shock
responds to local perturbations before conditions far downstream have
time to react.  Third, since $\Gamma^{-1}$ is used as a small
parameter, the pre-shock density fluctuations should not be so large
as to cause the shock to become sub-relativistic, \emph{i.e.} one
requires $\rho_0/\bar\rho_0\ll \bar\Gamma^{1/\lambda}$.  Finally,
transitions in density should not be so abrupt as to cause strong
reverse shocks, which would alter the forward shock dynamics.  Paper I
describes tests of ultra-relativistic GSD by comparison with exact
self-similar solutions (some of which it reproduces exactly) and with
numerical simulations.  The latter indicate, for example, that
eqs.~(\ref{eq:gamma}) and (\ref{eq:p}) are in error by only a few
percent for a $\bar\Gamma=10^2$ shock encountering overdensities as
large as $\rho_0/\bar\rho_0\lesssim 30$.

\subsection{Relativistic vorticity}\label{subsec:vort}

This subsection is independent of GSD.  We review the meaning of
enthalpy current and vorticity in ideal relativistic fluids,
especially those with the ultra-relativistic equation of state
$P=\rho/3$.

When the shock passes over a local density excess---considered, for
simplicity, in isolation from other inhomogeneities---the resulting
post-shock velocities are of two kinds.  First, since the shocked
clump is overpressured compared to its post-shock surroundings
[eq.~(\ref{eq:p})], it will expand and drive an outgoing pressure
wave. If the density contrast of the clump is small, then the wave is
essentially a linear disturbance from the start and travels at the
sound speed, $c/\sqrt{3}$, in the rest frame of the mean post-shock
flow; waves launched by large overdensities will be somewhat faster
and may steepen into secondary shocks, but whatever its strength, the
pressure wave rapidly departs its source. Overlapping pressure waves
launched by many distant clumps may contribute significant local
velocity perturbations, but, because of their oscillatory nature,
intuition suggests that these velocities will not secularly amplify
the magnetic field (except insofar as secondary shocks may contribute
to vorticity---see \S IV).  It would be interesting to test this
expectation in numerical simulations.

Unless the density excess is constant along the shock front, the
post-shock velocity field will also contain a vortical component,
whose strength is estimated below for an initially spherical
overdensity with a gaussian radial profile.  As shown in Paper I, the
equation of state $P=\rho/3$ allows some freedom in how one defines
relativistic vorticity, but to be useful in constraining the evolution
of the flow, the vorticity should be associated with a conservation
law such as Kelvin's Circulation Theorem,
\begin{equation}
  \label{eq:Kelvin}
  \frac{\ud}{\ud t}\oint\limits_C H_\mu \ud x^\mu =0,
\end{equation}
where $C$ is a closed contour comoving with the fluid four-velocity
$U^\mu$, and $H^\mu= h U^\mu$ for an appropriate thermodynamic
function $h$ (see below).  Equation (\ref{eq:Kelvin}) is equivalent to
\begin{equation}  \label{eq:vorteqn}
  \frac{\partial\bomega}{\partial t}-\grad\cross(\vv\cross \bomega)
=0\,,\qquad\bomega\equiv c\,\grad\cross\bH\,,
\end{equation}
where $v^i\equiv c \,U^i/U^0$ is the three-velocity of the fluid, and
$\bH$ is the spatial part of $H^\mu$.  Although formally identical to
the non-relativistic vorticity equation and written with
three-vectors, eq.~(\ref{eq:vorteqn}) is actually relativistically
covariant.

Because vorticity and circulation travel with the
local flow velocity and are nonoscillatory in the local fluid frame,
they have the potential to twist up and secularly amplify any magnetic
field frozen into the flow.  {\it A fundamental assumption of this
paper is that the magnetic energy will eventually reach equipartition
with the kinetic energy invested in the vortical part of the flow.}
This assumption would also be well worth testing numerically.

For a general equation of state $P=P(\rho,N)$, where $N$ is the proper number
density of conserved particles (\emph{e.g.}, baryons),
it is conventional to take the quantity $h$ in the relation $H^\mu=h U^\mu$
to be the enthalpy per particle: $h\equiv (\rho + P)/N$.  One assumes an ideal fluid,
so that the energy-momentum tensor is
\begin{equation}
  \label{eq:T}
T^{\mu\nu}\equiv(\rho+P)U^\mu U^\nu+P g^{\mu\nu}\,,
\end{equation}
($g^{\mu\nu}\to\mbox{diag}(-1,1,1,1)$ in Minkowski coordinates),
and the equations of motion are
\begin{equation}
  \label{eq:eoms}
T^{\mu\nu}_{~~~;\,\nu}=0\,\qquad\mbox{and}\quad(N U^\mu)_{;\,\mu}=0.
\end{equation}
The First Law of Thermodynamics
$dh= TdS+N^{-1}dP$ implies $h_{,\,\mu}=P_{,\,\mu}/N$ if the fluid is
isentropic; the first equation of motion in eq.~(\ref{eq:eoms})
can then be recast as
\begin{equation}
  \label{eq:eom}
U^\nu H^{\mu}_{~;\,\nu}=-h_{,\,\mu}\,,
\end{equation}
the ``curl'' of which implies eqs.~(\ref{eq:Kelvin}) and
(\ref{eq:vorteqn}) \citep[\emph{e.g.},][and Paper I]{Eshraghi03}.
These relations do not hold across shocks, of course, since there is
an entropy jump.  But if the pre-shock medium is inhomogeneous, then
even if the flow behind the shock is smooth, it will not be isentropic
in general; that is, $P/N^{4/3}$ (for an ultra-relativistic equation of state) will not be uniform.  
Therefore, the conventional choice of $h$ does not lead to a conserved circulation
under the circumstances contemplated in this paper.  Fortunately, as
pointed out in Paper I, if one defines $H^\mu$ using $h\propto
P^{1/4}$ instead of the true enthalpy, then eq.~(\ref{eq:eom})
always holds in smooth parts of the flow.  This is a consequence of
the equation of state $P=\rho/3$, which is ``barytropic'' if not
isentropic.  It is convenient to choose the constant of
proportionality so that $H^\mu$ reduces to the fluid four-velocity when
pressure is uniform.  Therefore, we replace the conventional enthalpy
current with
\begin{equation}
  \label{eq:Hdef}
  H^\mu\equiv\left(P/\bar P\right)^{1/4}U^\mu\,.
\end{equation}
With this choice, circulation is conserved [eqs.~(\ref{eq:Kelvin}) \&
(\ref{eq:vorteqn})] everywhere {\it except across shocks}.

\subsection{Vorticity production by shocks}\label{subsec:vortprod}

\begin{figure}[htbp]
  \centering 
\includegraphics[width=6in]{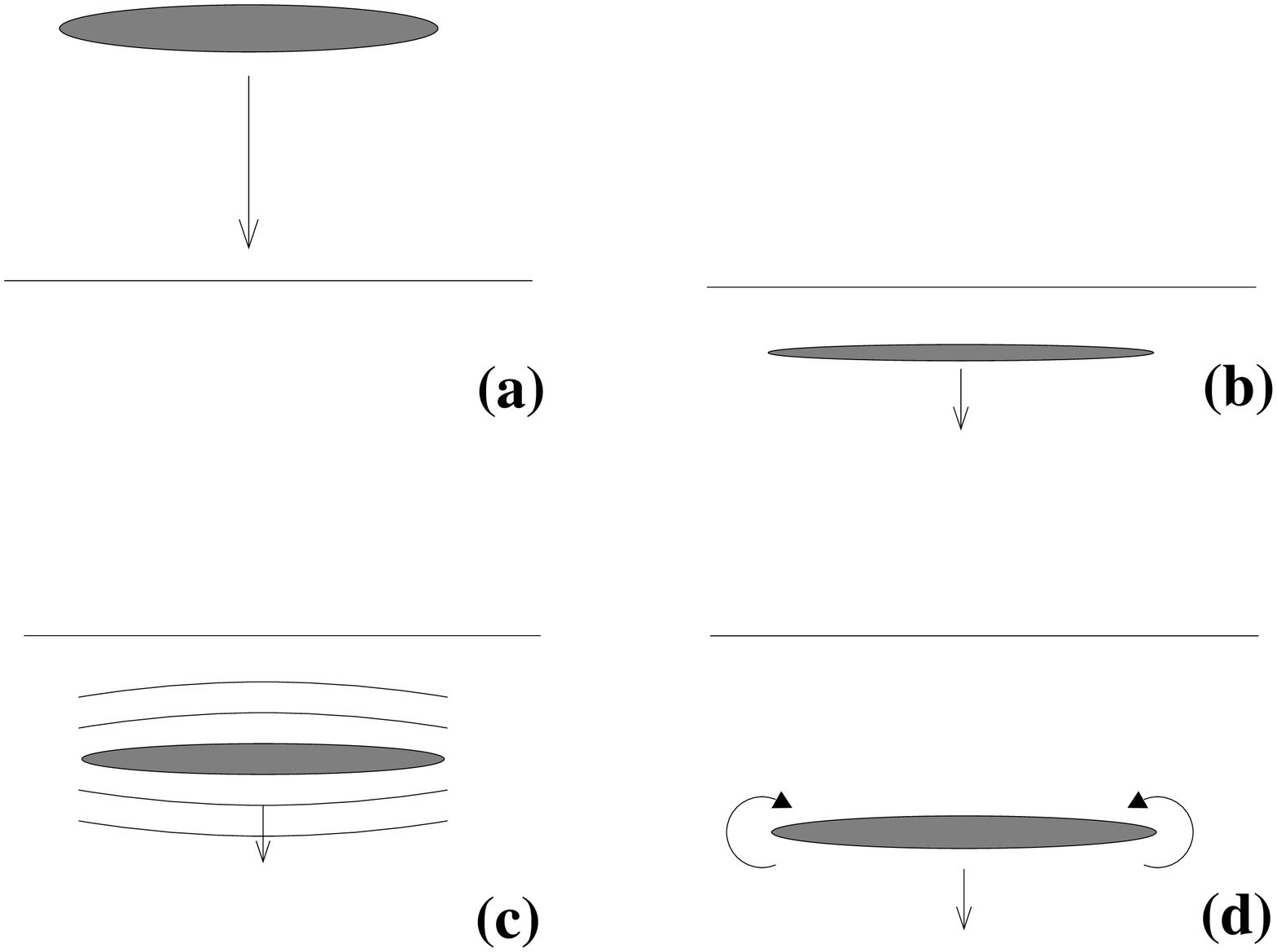}
\caption{History of a density clump overrun by an
  ultra-relativistic shock (mean Lorentz factor $\bar\Gamma$), viewed
  in the shock rest frame.  (a) The Lorentz-contracted (by a factor
  $\sim\bar\Gamma^{-1}$) spherical clump approaches the shock (thick
  line) with $v\approx c$.  (b) After the shock, the clump contracts
  further by $\times1/3$ (or $\times (2\sqrt{2})^{-1}$ in its rest
  frame) and moves downstream at $c/3$.  (c) The clump re-expands to
  reach pressure equilibrium and emits sound waves or weak shocks.
  (d) Vorticity created by the shock passage begins to roll up the clump.}
  \label{fig:cartoon} 
\end{figure}

The goal of this subsection is to use the one-dimensional GSD
approximation to derive eqs.~(\ref{eq:omegashock})-(\ref{eq:fdelta}), which relate the post-shock
vorticity to the pre-shock fractional overdensity
$\delta\equiv(\rho_0/\bar\rho_0)-1$.  Figure~\ref{fig:cartoon} illustrates
four stages in the interaction of an ultra-relativistic shock with a
density clump.

We begin by recalling some basic consequences of the shock jump
conditions that will be needed below.  Let $[Q]$ denote the
discontinuity in a fluid property $Q$ across the shock front.  In the
instantaneous local rest frame of the shock, where the outward unit normal to
the shock front is $\bn$, the jump conditions are $[T^{\mu j} n_j]=0$.
Using eq.~(\ref{eq:T}) for $T^{\mu\nu}$ (ideal fluid) and assuming that $P_0\ll\rho_0$
ahead of the shock and $P=\rho/3$ behind it, one finds that the
post-shock three-velocity of the fluid is $\vv\cdot\bn=-c/3$ in the
shock frame (hence subsonic, since the sound speed is $c/\sqrt{3}$).
The post-shock energy density is $\rho=2\Gamma^2\rho_0$, where $\Gamma\gg1$ is the local
Lorentz factor of the shock in the rest frame of the pre-shock fluid.
Similarly, conservation of particles implies $[NU^i n_i]=0$, whence
$N=2\sqrt{2}\Gamma N_0$.  The quantities $\rho$, $P$, and $N$ will
always denote proper values, meaning that they are defined in the
local fluid rest frame and are therefore Lorentz invariants by fiat, with
subscript ``0'' denoting a pre-shock value rather than a spacetime
index.

To facilitate Lorentz boosts between the pre-shock and post-shock or
shock frames, it is often convenient to use the rapidity parameter
$\tanh^{-1}(v/c)$, where $v$ is the three-velocity in the pre-shock
frame.  The relativistic addition of colinear three-velocities is
equivalent to addition of the corresponding rapidity parameters.  Thus,
for example, using $\Phi=\cosh^{-1}\Gamma\approx \ln(2\Gamma)$ for
the rapidity parameter of the shock and $\phi$ for the rapidity of the
post-shock fluid, it follows from the above that
$\tanh(\phi-\Phi)=-1/3$, whence $\phi=\Phi-\ln\sqrt{2}$.  The Lorentz
factor of the post-shock fluid relative to the pre-shock frame is
$\cosh\phi=\cosh(\Phi-\ln\sqrt{2})=\Gamma/\sqrt{2}+O(\Gamma^{-1})$.

Pre-shock clumps will typically have comparable longitudinal and
lateral dimensions (meaning: along and perpendicular to the \textit{mean}
direction of shock propagation) in their own rest frame.  In the shock
and post-shock frames, the clumps will be longitudinally contracted by
factors $\sim\Gamma^{-1}\ll 1$.  During the transit of the shock over
a clump, and even during the subsequent expansion of the shocked clump
as it comes to pressure equilibrium with the surrounding post-shock
fluid, there will not be enough time for signals (sound waves) to
communicate laterally from one end of the clump to the other.
Therefore, the interaction of the shock with the clump can be
calculated in a one-dimensional approximation, in which the area of
the shock is constant and the pre-shock mass density ($\rho_0/c^2$) is
stratified on planes parallel to the shock front.

Nevertheless, lateral density gradients do produce post-shock vorticity,
$\bomega$, even in our one-dimensional approximation.  First of all, the
longitudinal component $\bH_\parallel$ of the non-conventional enthalpy current defined in eq.~(\ref{eq:Hdef})
varies with lateral position behind the shock.  Secondly, since the shock
itself is delayed differently at different lateral positions, the
shock normal develops a lateral component, which leads to a small
lateral current, $\bH_\perp$. Although the magnitude of $\bH_\perp$ is $O(\Gamma^{-1})$ compared to $\bH_\parallel$ for a given clump amplitude, its
longitudinal derivative makes a contribution to $\bomega$ that is comparable to the lateral derivative of $\bH_\parallel$, as a consequence of Lorentz contraction.

Let $\delta(\br)$ be the density contrast of the clump:
$\rho_0(\br)=\bar\rho_0[1+\delta(\br)]$, where $\bar\rho_0$ is the
mean pre-shock value. The shock propagates along $z$ on average; let us choose a cartesian coordinate system $x, y, z_0$ in the pre-shock rest frame, where the subscript ``0'' is used to distinguish between the post-shock ($z$) and pre-shock ($z_0$) longitudinal coordinates. In our one-dimensional GSD approximation, it follows from eq.~(\ref{eq:gamma}) that
\begin{equation}
  \label{eq:Gammadelta}
  \Gamma(\br)=\bar\Gamma[1+\delta(\br)]^{-\lambda}\,,
\end{equation}
if $\bar\Gamma\gg1$ is the value for a smooth pre-shock medium. As discussed at the beginning of this section, it will be assumed that $|\delta|^\lambda\ll\bar\Gamma$ so that
$\Gamma\gg1$ at all times; since $\lambda\approx 0.232$ is fairly
small, rather hefty density contrasts can be compatible with this
condition. 

First of all, let us focus on the contribution of $\bH_\perp$ to the post-shock vorticity. Let $\tau(\br)$ be the time in the pre-shock frame at
which the shock reaches pre-shock position $\br$:
\begin{equation}
  \label{eq:tauofx}
  c\,\tau(x,y,z_0)\approx z_0+\frac{1}{2}\bar\Gamma^{-2}\int\limits^{z_0} 
[1+\delta(x,y,z_0')]^{2\lambda}\,\ud z_0'\equiv z_0+\frac{1}{2}\bar\Gamma^{-2} I(x, y, z_0).
\end{equation}
where we have assumed that the typical size of a density clump is
small compared to the characteristic lengthscale for variations in
$\bar{\Gamma}$.  The shock surface at time $t$ is determined
implicitly by $\tau(\br)=t$, so that the shock normal is
$\bn=|\grad_0\tau|^{-1}\grad_0\tau$ (the subscript reminds that derivatives are taken ahead of the shock).  To leading order in
$\bar\Gamma^{-1}$, the lateral components of the normal are therefore
\begin{equation}\label{eq:n}
  \bn_\perp\approx \lambda\bar\Gamma^{-2}\grad_{\perp} I= 
\lambda\bar\Gamma^{-2}\int\limits^{z_0} 
(1+\delta')^{2\lambda-1}(\grad_\perp\delta')\,\ud z_0'.
\end{equation}
Here $\delta'$ is shorthand for $\delta(x,y,z_0')$ and $\grad_{\perp}\equiv(\partial/\partial x, \partial/\partial y, 0)$ is the same in the pre-shock and post-shock reference frames.
The lateral part of the post-shock fluid 4-velocity is
$\b{U}_\perp\approx\bn_\perp U_\parallel\approx\bn_\perp\Gamma/\sqrt{2}$.
With use of eqs.~(\ref{eq:Hdef}) and (\ref{eq:Gammadelta}),
the post-shock lateral enthalpy current becomes
\begin{equation}
  \b{H}_\perp\approx 2^{-1/2}\lambda\bar\Gamma^{-1}(1+\delta)^{(1-6\lambda)/4}
\grad_{\perp} I.
\end{equation}
Of course the lateral components of the enthalpy current take the same
values in the mean post-shock and pre-shock frames since these frames
differ by a longitudinal boost.  However, in order to compute the
vorticity in the mean post-shock fluid frame, we should remember that
the longitudinal derivative in this reference system
$\grad_\parallel$ is related to the corresponding derivative $\grad_{0,\parallel}$ in
the pre-shock frame by $\grad_\parallel=\cosh\bar\phi\,\grad_{0,\parallel}$
because of Lorentz contraction.  Thus the contribution of
$\b{H}_\perp$ to the post-shock vorticity is
\begin{eqnarray}
  \label{eq:parperp}
    \grad_{\parallel}\cross\b{H}_\perp&\approx&
\frac{1}{2}\lambda(1+\delta)^{(2\lambda-3)/4}(\bar\bn\cross\grad\delta)\nonumber\\
&+&
\frac{\lambda(1-6\lambda)}{8}(1+\delta)^{-3(1+2\lambda)/4}
\frac{\partial\delta}{\partial z_0}(\bar\bn\cross\grad I)
\end{eqnarray}
where $\bar\bn=\b{e}_z$ is the mean shock normal and clearly the differential operator $\bar\bn\cross\grad$ is the same in the pre-shock and post-shock mean rest frames.

A comparable contribution to the post-shock vorticity comes from $\bH_\parallel$. The longitudinal component of the post-shock 4-velocity is
$\b{U}_{||}=\sinh(\phi-\bar\phi)\,\be_z$ when measured in the mean post-shock
frame.  Since $\phi-\bar\phi=\Phi-\bar\Phi=-\lambda \ln(1+\delta)$,
\begin{equation}
  \label{eq:Upar}
  \b{U}_\parallel=-\frac{1}{2}\left[(1+\delta)^\lambda-(1+\delta)^{-\lambda}\right]\,\be_z
\end{equation}
in the mean post-shock frame.  This tends to be quite subluminal:
for unit overdensity ($\delta=1$), for example,
$\b{U}_\parallel\approx -0.1615\,\b{e}_z$, and the corresponding fluid
Lorentz factor is $\approx 1.013$. The longitudinal post-shock
enthalpy current is $\b{H}_\parallel=(P/\bar P)^{1/4}\b{U}_\parallel=
(1+\delta)^{(1-2\lambda)/4}\b{U}_\parallel$.  Taking the lateral
derivative of this and combining with eq.~(\ref{eq:parperp}) and the
definition of $I$ in eq.~(\ref{eq:tauofx}), we can write the total
post-shock vorticity as
\begin{eqnarray}
  \label{eq:omegashock}
  \b{\omega}&\approx&c\,(\bar\bn\cross\grad\delta)f(\delta)\nonumber\\
&+&c\,\frac{\lambda(1-6\lambda)}{8}(1+\delta)^{-3(1+2\lambda)/4}
\frac{\partial\delta}{\partial z_0}
\int\limits^{z_0} (1+\delta')^{2\lambda-1}(\bar\bn\cross\grad\delta')\,\ud z_0'\;,
 \end{eqnarray}
 where the coordinate system is in the pre-shock rest frame and the
 derivatives are taken ahead of the shock. We have
 introduced the function
  \begin{equation}\label{eq:fdelta}
f(\delta)\equiv(1+\delta)^{-(2\lambda+3)/4}\left[
\frac{6\lambda+1}{8}(1+\delta)^\lambda
+\frac{6\lambda-1}{8}(1+\delta)^{-\lambda}\right].
\end{equation}
For $|\delta|\ll 1$, $f(\delta)\approx3\lambda/2\approx0.348$.  It is
interesting that in the total post-shock vorticity there are no
surviving factors of $\bar\Gamma$.

\subsection{Vortical energy}\label{subsec:en}

We have not been able to formulate a rigorous relativistically
covariant way of dividing the energy of the post-shock flow into
vortical and non-vortical parts.  In a non-relativistic
flow, however, this would be straightforward.  One would divide the
three-velocity field into potential and solenoidal parts,
\begin{equation}
  \label{eq:dec}
  \vv=\grad\psi + \grad\cross\bA\,,
\end{equation}
and then define the vortical energy by
\begin{equation}\label{eq:Evortnr}
  E_{\rm vort}\equiv \frac{1}{2}\int\rho_m\left|\grad\cross\bA\right|^2\,\ud^3x\,,
\end{equation}
where $\rho_m$ is the non-relativistic mass density.  To make the
decomposition (\ref{eq:dec}) unique, some mild additional restrictions
are necessary: for example, that the region of interest is simply
connected, and that $\psi$ and $\bA$ have some specified behavior on
the boundary or at infinity. Then one can impose $\grad\cdot\bA=0$ and
solve for $\bA$ from
\begin{equation}
  \label{eq:Asol}
  \nabla^2\bA=-\bomega\,,
\end{equation}
with $\bomega\equiv\grad\cross\vv$, using an appropriate Green's
function to invert $\nabla^2$.  The nonvortical part follows similarly
from $\nabla^2\psi=\grad\cdot\vv$.

It isn't clear how to proceed in the relativistic case because the
coordinate energy density $T^{00}$ isn't simply quadratic in $\vv$, in
general.  However, for the applications we have in mind, the vortical
motions are plausibly subsonic in the mean post-shock frame (as long
as $\delta$ is not too large), and therefore only mildly relativistic,
$(v/c)^2\lesssim 1/3$.  Therefore, we will use the decomposition
(\ref{eq:dec}), except for three changes: (i) $c\,\bH$ replaces $\vv$,
and therefore $\bomega\to c\,\grad\cross\bH$ in eq.~(\ref{eq:Asol});
(ii) $\rho_m$, the non-relativistic mass density, is replaced by
$\rho/c^2$, where $\rho$ is the proper internal energy density; (iii)
the factor $1/2$ in eq.~(\ref{eq:dec}) is replaced by $4/3$.  The
reason for (iii) is that if $v/c\ll 1$ and $P=\rho/3$, then
$T^{00}\approx\rho+(4/3)\,\rho \,(v/c)^2$, and we assume that $E_{\rm
  vort}$ derives from the second term.\footnote{Why $4/3$ rather than
  $1/2$, as one might expect in the non-relativistic limit?  The
  answer is that the rest-mass density of non-relativistic fluid
  mechanics is actually $\rho_m=\gamma N m$, where $m$ is the mass per
  particle and $\gamma=U^0$, whereas the proper energy density is
  $\rho= Nm(c^2+u)$, where $mu$ is the internal energy per particle.
  Thus, for a general equation of state,
  $T^{00}=\gamma^2\rho+(\gamma^2-1)P=\gamma\rho_m(c^2+u)+(\gamma^2-1)P$.
  For $(v/c)^2\ll 1$, this reduces to $T^{00}\approx\rho_m
  c^2+\rho_m(v^2/2+u)+[(v/c)^2(\rho_m u/2+P)]$. The middle term is now
  recognizable as the non-relativistic kinetic-plus-internal energy
  density.  The term in square brackets would normally be negligible
  for a cold fluid because $u\sim P/\rho_m\ll c^2$; however, for an
  ultra-relativistic ideal fluid $u\approx 3P\gamma/\rho_m\gg c^2$ and
  the result is $T^{00}\approx\rho_m c^2+\rho_m v^2/2 +3P+4P(v/c)^2$,
  of which the first two terms are now negligible.}

A pre-shock density clump will experience, after the transit across
the shock front, a contraction of its longitudinal size by a factor
$\alpha^{-1}$, where 
$\alpha\equiv N/N_0\approx 2\sqrt{2}\bar\Gamma\gg1$.  What was
an approximately spherical clump becomes a pancake, with all of
its associated vorticity in or near the interior.  
We adopt post-shock cylindrical coordinates $R,\theta,z$ with
the $z$ axis parallel to the mean direction of propagation of the
shock.  Taking the origin at the center of the clump, assuming
the clump to be axisymmetric, and recalling that the vorticity
computed in eq.~(\ref{eq:omegashock}) has been expressed in pre-shock
coordinates, we may write
$\b{\omega}=\omega(R, z_0)\,\be_{\theta}=\omega(R,
\alpha z)\,\be_{\theta}\approx \sigma(R)\delta_{\msc{d}}(z)\be_\theta$,
where $\delta_{\msc{d}}(z)$ is the Dirac delta function, and
\begin{equation}\label{eq:proj}
\sigma(R)\equiv\int\limits^{+\infty}_{-\infty}\omega(R, \alpha z)\,\ud z=
\frac{1}{\alpha}\int\limits^{+\infty}_{-\infty}\omega(R,  z_0)\,\ud z_0
\end{equation}
will be referred to as the ``projected vorticity''.

Just after the shock passage, a clump will be at higher pressure
than its surroundings, as already discussed, and will therefore expand
until it reaches approximate pressure equilibrium.  Since the
sound-crossing time in the longitudinal direction will be much less
than that in lateral directions, equilibrium will be reached with only
a small fractional change in the lateral size.  The fractional change
in the longitudinal size will be of order unity, but for
$\bar\Gamma\gg 1$ and clump overdensities $\delta\ll
\bar\Gamma^{\gamma_{\rm{ad}}/(1-2\lambda)}$ ($\gamma_{\rm{ad}}=2$ is
the adiabatic index appropriate for the one-dimensional expansion of
an ultra-relativistic fluid) the clump will remain highly flattened
even after equilibration.  Since the vortex lines lie parallel to the
clump midplane (\emph{i.e.}, in the direction $\be_\theta$) and are
``frozen'' into the clump (as a consequence of Kelvin's Circulation
Theorem), the expansion will have little effect on the projected
vorticity in eq.~(\ref{eq:proj}).  We may therefore solve
eq.~(\ref{eq:Asol}) in the approximation that the vorticity is
confined to a thin sheet with the projected vorticity computed in
eq.~(\ref{eq:proj}). The problem is mathematically equivalent to
finding the vector potential due to an axisymmetric current sheet
carrying a toroidal current:
\begin{equation}
\b A(R, z)= A(R, z)\b e_{\theta}=
\frac{1}{2}\int\limits^{+\infty}_0\tilde\sigma(k)\,J_1(kR)\,e^{-k|z|}\,\ud k\;
\b e_{\theta}
\end{equation}
where $J_1(kR)$ is the Bessel function of order $1$ and
$\tilde{\sigma}(k)$ is the Hankel transform of the projected vorticity
$\sigma(R)$:
\begin{equation}
\tilde\sigma(k)\equiv\int\limits_0^{+\infty}\sigma(R)\,J_1(kR)\,R\, \ud R\;.
\end{equation}
As soon as the clump is at the same pressure as its surroundings, if the vortical motions are sub-relativistic we can use the decomposition (\ref{eq:dec}) with $\vv_{\rm vort}\to c\,\bH_{\rm vort}=\grad\cross\bA$. Moreover, pressure equilibrium between the clump and the average post-shock medium implies
\footnote{In the mean post-shock rest frame, if $T^{\mu\nu}$ is the energy-momentum tensor of the fluid in turbulent motion and $\bar T^{\mu\nu}$ for the mean fluid (whose four-velocity has spatial components $\bar{U}_i=0$ in this reference frame), pressure equilibrium requires $T^{ij}=\bar T^{ij}$; under the assumption of spatial statistical isotropy for the turbulent motions, this implies $T^{kk}=3\bar P$, which yields, for an ultra-relativistic fluid ($T^{\mu\mu}=\bar T^{\mu \mu}=0$), the equality $T^{00}=\bar\rho$; then, since $T^{00}\approx\rho+(4/3)\,\rho \,(v/c)^2$, if the vortical motions are sub-relativistic we obtain $\rho\approx\bar\rho$.}
 that the clump proper energy density $\rho$ is approximately equal (neglecting terms of order $O(v^2/c^2)$) to the mean post-shock value $\bar\rho$. Then the vortical energy of a single clump becomes
\begin{equation}\label{eq:Evort2}
E_{\rm{vort}}\equiv\frac{4}{3}\bar\rho\int\frac{|\grad\cross\b A|^2}{c^2}\ud^3x=
-\frac{4}{3}\bar\rho\int\frac{\bA\cdot\grad^2\b A}{c^2}\;\ud^3x\;
=\frac{4}{3}\bar\rho\int\frac{\bA\cdot\bomega}{c^2}\;\ud^3x\;.
\end{equation}
In accordance with the discussion following (\ref{eq:Asol}), we have
replaced the factor $1/2$ in (\ref{eq:Evortnr}) with $4/3$.  Also, we
have used integration by parts to replace the integral over all space
with an integral over the source of vorticity only.  This is
particularly convenient when the source is represented by a vortex
sheet, because we may express the energy as an integral in $k$ space:
\begin{equation}\label{eq:vortfrac2}
E_{\rm{vort}}=\frac{8\pi}{3}\bar\rho\int\limits_0^{+\infty}\frac{ A(R, 0)\,\sigma(R)}{c^2}\,R\,\ud R=
\frac{4\pi}{3}\bar\rho\int\limits_0^{+\infty}\frac{\tilde{\sigma}^2(k)}{c^2}\,\ud k\;. 
\end{equation}

We shall assume that the vortical energies of different clumps can
simply be added.  This is justified if the clumps are well
separated compared to their larger (\emph{i.e.}, lateral) dimensions.
Then if the number density of clumps in the pre-shock frame is
$N_{\rm{c,0}}$ and all the clumps have the same axisymmetric density
profile, the vortical energy density in the average post-shock frame
is $\rho_{\rm{vort}}=\alpha\, N_{\rm{c,0}} E_{\rm{vort}}$, and
the vortical energy fraction becomes
\begin{equation}\label{eq:vortfrac}
\epsilon_{\rm{vort}}\equiv\frac{\rho_{\mr{vort}}}{\bar\rho}=
\frac{4\pi}{3}\alpha\, N_{\rm c,0}\int\limits_0^{+\infty}\frac{\tilde{\sigma}^2(k)}{c^2}\,\ud k\;. 
\end{equation}
Recall from eq.~(\ref{eq:proj}) that the projected vorticity is
proportional to $\alpha^{-1}=(2\sqrt{2}\bar\Gamma)^{-1}$.  Since this
factor is squared in computing the vortical energy, it follows
from eq.~(\ref{eq:vortfrac}) that $\epsilon_{\rm
  vort}\propto\bar\Gamma^{-1}$ for a fixed pre-shock density field.
This scaling is perhaps the most important conclusion of our analysis
up to this point.

\section{The circumburst medium}\label{sec:medium}

Observations support the idea that long-duration GRBs are associated
with the deaths of massive Wolf-Rayet (WR) stars, presumably arising
from their core-collapse \citep[][and references therein]{Woosley_Bloom06}.
Then 
the circumburst environment is determined by the star's mass-loss history.  At
the onset of the WR phase, the WR stellar wind is expected to expand
with a typical velocity $v_{\msc{wr}}\approx2000\unit{km\,s^{-1}}$ inside the
pre-existing slower wind emitted during the red supergiant (RSG)
phase, whose characteristic speed is
$v_{\msc{rsg}}\approx20\unit{km\,s^{-1}}$. The winds from massive RSGs are
characterized by a mass-loss rate $\dot{M}_{\msc{rsg}}$ between
$10^{-6}$ and $10^{-4}\unit{M_\odot yr^{-1}}$
\citep{Chevalier_Fransson_Nymark06}, while the mass-loss rates $\dot{M}_{\msc{wr}}$ of WR stars are between $10^{-5}$ and
$10^{-4}\unit{M_{\odot}yr^{-1}}$ \citep{Crowther06}.  Several solar
masses are shed by the star during these evolutionary
phases.  The mass equivalent to the energy of a GRB, on
the other hand, is only $\approx0.06 \,E_{\rm iso,53} \unit{M_\sun}$.  Therefore, the
GRB forward shock is expected to become non-relativistic long before it
escapes the wind to encounter the interstellar medium.  There are at
least four regions of the wind that are relevant to the relativistic
phase of the afterglow \citep{Ramirez-Ruiz_etal05}: from the inside out,
these are an expanding WR wind ($\bar\rho_0/c^2=\dot M_{\msc{wr}}/4\pi v_{\msc{wr}} r^{2}$, where $r$ is the distance from the star),
the shocked WR wind ($\bar\rho_0\approx\mbox{constant}$), the shocked RSG wind
($\bar\rho_0\approx\mbox{constant}$), and a freely expanding RSG wind
($\bar\rho_0/c^2=\dot M_{\msc{wr}}/4\pi v_{\msc{rsg}}r^2$).  Beyond these lie another
shocked part of the RSG wind, the shocked ISM, and finally the unshocked ISM.

Density inhomogeneities in such a stratified structure could be
produced by several processes.  First of all, the acceleration region
of the WR wind, which extends to a few times the stellar radius, is known
to be clumpy.  Emission-line data indicate accelerating ``blobs''
\citep{Moffat_etal88} with density filling factors $f\approx0.05-0.25$
\citep[and references therein]{Crowther06}.  These must occur where
the continuum is optically thin and must be large enough transversely
to cover an appreciable part ($\gtrsim 10\%$) of the stellar
photosphere.  The clumpiness of this region is in accord with theory,
since the line-driving mechanism of \citet{CAK75}, which explains the
gross properties of WR and main-sequence O-star winds rather
satisfactorily, is known to be unstable both radially and nonradially
on scales larger than the ``Sobolev length'' $l_{\mr{Sob}}\sim
r\,v_{\mr{th}}/v_{\msc{wr}}\approx 10^{-2} r$ \citep[and references
therein]{Dessart_Owocki05}, where $v_{\mr{th}}\approx20\unit{km\,s^{-1}}$ is the thermal velocity in the wind.  Clump dimensions at least as small as
$\sim 10^{10}\unit{cm}$ with density contrasts $\sim f^{-1}\sim 10$ are
therefore expected.  These clumps may dissolve beyond the acceleration
region ($r\gg 10^{12}\unit{cm}$) if the optically thin wind maintains a
uniform temperature comparable to the color temperature of the star;
the clumps are then at higher density than their surroundings and
will expand on their sound-crossing time.  After crossing the reverse
shock, the wind reaches temperatures $\sim 10^8\unit{K}$, a regime
that is thermally unstable \citep{field_65}.  Beyond the
contact discontinuity, the shocked RSG wind lies at
$T\sim10^6\unit{K}$, which is even more unstable.  Thermal
instability may give rise to new clumps, whose minimum size is
controlled by thermal conduction and is therefore very uncertain
because the conduction rate is probably sensitive to magnetic field.
Furthermore, the contact discontinuity is subject to Rayleigh-Taylor
instability as the shocked RSG wind is accelerated by the less dense
WR wind; however, \citet{garcia-segura_franco96} found that clump formation is not
efficient for the case of pure Rayleigh-Taylor instabilities in thick
shells and concluded that a necessary condition for clump development
to occur is that the shocked shell be thin enough to allow
Vishniac-like instabilities \citep{vishniac_83}.
\citet{Ramirez-Ruiz_etal05} point out that rapid cooling of the
shocked RSG wind is indeed likely to lead to a thin shell.  Clump
development in the shell is controlled by the Kelvin-Helmholtz damping
of the coupling between the Rayleigh-Taylor and Vishniac
instabilities, while the lifetime of small dense clumps, once formed,
probably again depends upon conduction.

In short, there are ample reasons to expect a strongly inhomogeneous
medium ahead of the GRB forward shock, but the sizes of the smallest clumps
are uncertain.  We make no attempt to estimate these sizes but simply
state below the properties (lengthscale, density contrast and number density) that the
clumps would have to have in order to produce post-shock turbulence
capable of amplifying the magnetic field sufficiently.

It is of course essential for our proposed mechanism that a seed field
exist in the pre-shock medium.  Magnetic and thermal energy densities
are generally comparable in the ISM, but much less is known about the
magnetic component of the winds of early-type stars.  Since such winds
are believed to be driven by radiative rather than centrifugal forces,
as in late-type main-sequence stars, and since such stars do not
possess surface convection zones, the magnetic field might be very
small $\emph{a priori}$.  However, magnetic effects have been invoked
to explain X-ray emission and anisotropy in the winds of early-type
stars \citep{wolf_etal99, ud-doula_owocki02, schulz_etal03}.  Furthermore, since the amplification of a
weak (kinematic) magnetic energy density by turbulence proceeds
exponentially on the timescale of the energy-bearing eddies
\citep[and references therein]{schekochihin_etal02}, the number $N_{\mr{eddy}}$ of eddy-turnover times required to
reach saturation is only logarithmically dependent on the strength of
the seed field.  Therefore, in the estimates of the required turnover
time made below, we shall simply assume that the magnetic energy fraction $\epsilon_{\msc{b}}$ in the pre-shock medium is comparable to
that in the ISM, $\sim 10^{-8}$, so that the number of eddy rotations necessary to explain the value of $\epsilon_{\msc{b}}$ inferred from afterglow observations is of order $N_{\mr{eddy}}\sim10$.

\section{Results}\label{sec:results}

The formalism described in \S II can be used to predict the eddy-turnover time and vortical energy fraction of the turbulent motions produced by an ultra-relativistic shock sweeping-up a clumpy medium; if the pre-shock average (\emph{i.e.}, smoothed over clumps) density profile is known, a comparison with afterglow models will then let us constrain the lengthscale, overdensity and volume filling factor of the circumburst inhomogeneities. Unfortunately, observations have not yet clearly determined the density profile ahead of the GRB forward shock \citep{ chevalier_li00, panaitescu_kumar01, panaitescu_kumar02, chevalier_etal04}, so that we will consider both free winds (energy density $\bar{\rho}_0\propto r^{-2}$) and shocked uniform winds ($\bar\rho_0\approx\mbox{constant}$) as possible circumburst media, keeping in mind that, as described by \citet{Ramirez-Ruiz_etal05} and outlined in \S III, the actual surrounding environment is much more complex.

Assuming that the blast wave is adiabatic and effectively spherical and that $E_{\mr{iso}}=10^{53}E_{\mr{iso, 53}}\unit{erg}$ is its isotropic equivalent energy, as derived from the gamma-ray output, we can compute the deceleration radius of the GRB forward shock when about half of the initial energy has been transferred to the shocked matter; for an initial Lorentz factor $\Gamma_0=10^2\Gamma_{0,2}$, the typical swept-up mass where this happens is
\begin{equation}
M_{\mr{dec}}=\frac{E_{\mr{iso}}}{\Gamma_0^2\, c^2}\approx5.6\times10^{-6}\,E_{\mr{iso, 53}}\,\Gamma^{-2}_{0,2}\,\unit{M_\odot}\,.
\end{equation}
For the two circumburst density profiles mentioned above, the deceleration radius of the shock is then 
\begin{equation}
\label{eq:rdec} 
r_{\rm dec} = \left\{ \begin{array}{ll} 
\frac{M_{\mr{dec}}\, v_{\msc{wr}}}{\dot{M}_{\msc{wr}}}\approx3.5\times10^{15}\,E_{\mr{iso, 53}}\,v_{\msc{wr},8.3}\,\dot{M}_{\msc{wr}, -5}^{-1}\,\Gamma^{-2}_{0,2}\,\unit{cm}
 & \textrm{for $\bar{\rho}_0\propto r^{-2}$}\\ 
 \left(\frac{3\,M_{\rm dec}}{4\pi m_p\, n_{\msc{ism}}}\right)^{1/3}
\approx1.2\times10^{17}\,(E_{\mr{iso, 53}}/n_{\msc{ism},0})^{1/3}\,\Gamma^{-2/3}_{0,2}\,\unit{cm}
& \textrm{for $\bar\rho_0\approx\mbox{const.}$} 
\end{array} \right. 
\end{equation}
where in the first case we have chosen typical WR wind parameters ($\dot{M}_{\msc{wr}}=10^{-5}\dot{M}_{\msc{wr}, -5}\unit{M_{\odot}yr^{-1}}$ and $v_{\msc{wr}}=2000\,v_{\msc{wr},8.3}\unit{km\,s^{-1}}$), whereas in the second case a baryon number density comparable to the ISM value ($n_{\msc{ism}}=n_{\msc{ism},0}\unit{cm^{-3}}$) has been assumed. This corresponds to a deceleration time in the post-shock rest frame
\begin{equation}
\label{eq:tdec} 
t_{\rm dec} =\sqrt{2}\frac{r_{\rm dec}}{\Gamma_0\,c}\approx\left\{ \begin{array}{ll} 
1.7\times10^3\,E_{\mr{iso, 53}}\,v_{\msc{wr},8.3}\,\dot{M}_{\msc{wr}, -5}^{-1}\,\Gamma^{-3}_{0,2}\,\unit{s} & \textrm{for $\bar{\rho}_0\propto r^{-2}$}\\ 
5.7\times10^4\,(E_{\mr{iso, 53}}/n_{\msc{ism},0})^{1/3}\,\Gamma^{-5/3}_{0,2}\,
\unit{s}& \textrm{for $\bar\rho_0\approx\mbox{const.}$} 
\end{array} \right. 
\end{equation}
since the Lorentz factor of the post-shock material in the pre-shock rest frame is $\Gamma_0/\sqrt{2}$. 

\subsection{Clump properties}

We consider a pre-shock medium with number density $N_{\rm{c,0}}$ of identical clumps. Each clump is characterized by a gaussian overdensity profile with central density contrast $\delta_{\rm{max}}$ and typical size $L$; if the origin of the axes is in the center of the clump and $R$, $z_0$ and $\theta$ are pre-shock cylindrical coordinates, we choose a clump energy density profile
\begin{equation}\label{eq:overdens}
\rho_0(R,z_0)=\rho_{\rm{ext}}[1+\delta_{\rm{max}}e^{-(R^2+z_0^2)/L^2}]\;,
\end{equation}
where $\rho_{\rm{ext}}$ is the energy density of the inter-clump homogeneous medium. In order to use the formalism introduced in \S II---where the density contrast $\delta$ was defined with respect to the mean pre-shock energy density $\bar\rho_0$ (averaged over clumps), whereas here the inter-clump density $\rho_{\mr{ext}}$ has been used---we should set
\begin{equation}
\delta(R, z_0)=\frac{\rho_{\rm{ext}}}{\bar\rho_0}[1+\delta_{\rm{max}}e^{-(R^2+z_0^2)/L^2}]-1\;;
\end{equation}
however, the ratio between $\rho_{\rm{ext}}$ and $\bar{\rho}_0$ can be easily computed for the density profile in eq.~(\ref{eq:overdens}):
\begin{equation}
\frac{\rho_{\rm{ext}}}{\bar{\rho}_0}=\frac{1}{1+\pi^{3/2}\delta_{\rm{max}}N_{\rm{c,0}} L^3}\approx1\label{rhoext}
\end{equation}
provided that $\delta_{\rm{max}}N_{\rm{c,0}} L^3\lesssim1$. In this section, we will always assume $\rho_{\rm{ext}}/\bar{\rho}_0\approx1$, so that the formulae in \S II can be used with a pre-shock overdensity profile
\begin{equation}\label{eq:density}
\delta(R, z_0)=\delta_{\rm{max}}e^{-(R^2+z_0^2)/L^2}\;.
\end{equation}

\begin{figure}[htbp]
  \centering 
\includegraphics[width=5in]{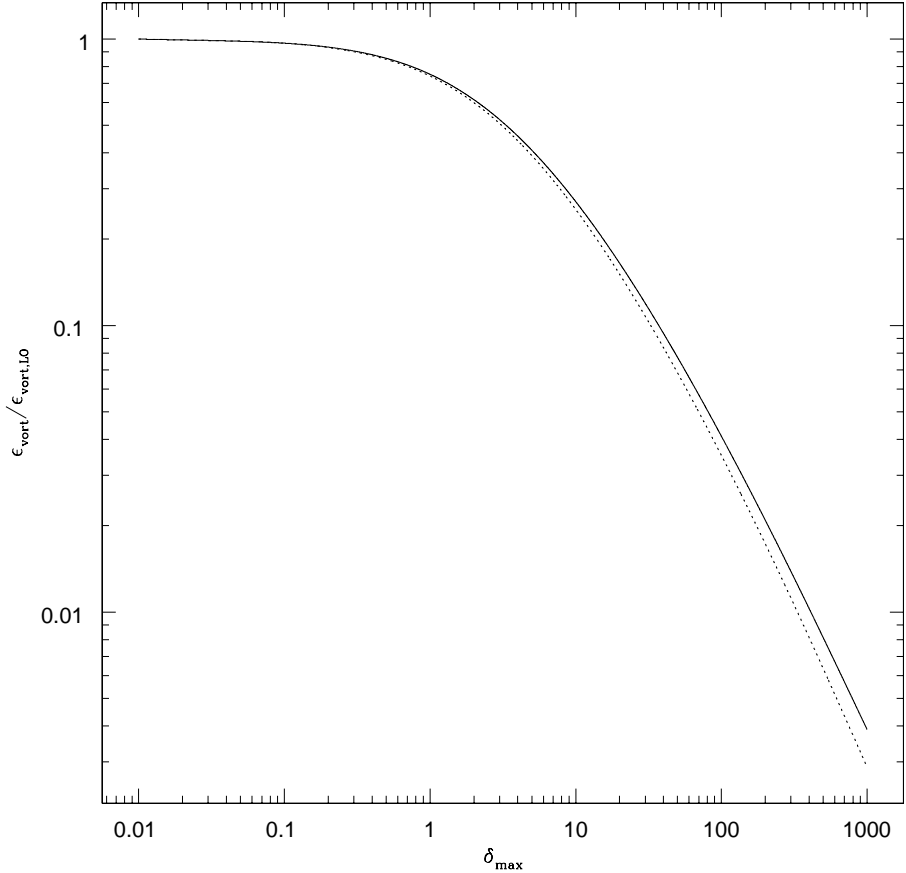}
\caption{
  Ratio between the exact numerical calculation of the vortical energy
  fraction ($\epsilon_{\mr{vort}}$) and its analytical approximation
  ($\epsilon_{\mr{vort},\msc{lo}}$ in eq.~(\ref{eq:energylimit}),
  keeping only the leading order in the central overdensity
  $\delta_{\mr{max}}$ within the vorticity $\bomega$) as a function of
  $\delta_{\mr{max}}$; deviations from the anaytical calculation are
  significant only for $\delta_{\mr{max}}\gtrsim1$. \textit{Dotted
    line}: Taking into account, in the numerical calculation, only the
  non-integral contribution to the vorticity (\emph{i.e.}, neglecting
  the second row in eq.~(\ref{eq:omegashock})). \textit{Solid line}:
  Full numerical calculation, including both the integral and
  non-integral terms in eq.~(\ref{eq:omegashock}); the non-integral
  contribution is found to be dominant even for large overdensities.
}
\label{fig:plot1} 
\end{figure}

For small density contrasts, we can perform analytical calculations
taking into account just the leading (first) order term in $\delta$
within eq.~(\ref{eq:omegashock}); the corresponding vorticity will be
referred to as the ``leading-order vorticity'' $\bomega_{\msc{lo}}$,
and the Hankel transform of its projected vorticity will be used in
eq.~(\ref{eq:vortfrac}) to analytically compute the leading (second)
order in $\delta_{\rm max}$ of the vortical energy fraction
($\epsilon_{\rm{vort}, \msc{lo}}$). This approximation is certainly
well justified for $\delta_{\mr{max}}\ll1$, but Figure~\ref{fig:plot1}
shows that the full numerical computation for $\epsilon_{\mr{vort}}$
is still in reasonable agreement with our analytical approximation
$\epsilon_{\rm{vort}, \msc{lo}}$ even for $\delta_{\mr{max}}\sim1$.
For larger central overdensities a numerical calculation is required,
and we could fit the numerical results with a fitting function
\begin{equation}
 \label{eq:fit}
 \epsilon_{\mr{vort}}=\epsilon_{\rm{vort}, \msc{lo}} \frac{1}{1+c_1\,(\delta_{\mr{max}})^{c_2}}\;
\end{equation}
($c_1$ and $c_2$ are fitting parameters), so that for small overdensities ($\delta_{\mr{max}}\ll1$) we recover the result of the analytical computation.

\subsection{Eddy-turnover time}\label{subsec:turnover}

Vorticity embedded in clumps by the passage of the GRB forward shock
can be responsible for the magnetic fraction inferred from afterglow
models only if vortical motions are fast and energetic enough to
amplify the field up to the observed value before the shock
deceleration time, when adiabatic expansion would significantly reduce
the particle energies available for the afterglow emission.

In the leading-order approximation described above, an estimate of the
eddy-turnover time for the density contrast in eq.~(\ref{eq:density})
is
\begin{equation}
t_{\rm{eddy}}=\frac{1}{|\b\omega_{\msc{lo}}(L, 0)|}\approx
6.5\times10^2\,L_{13}\frac{2}{\delta_{\max}}\unit s
\end{equation}
where $L=10^{13}L_{\rm{13}}\unit{cm}$;
the reference value chosen for the central overdensity
$\delta_{\rm{max}}$ is in agreement with observations (see \S III) and
reasonably satisfies the requirements for the leading-order
approximation (see Figure~\ref{fig:plot1}). If
$N_{\mr{eddy}}=10\,N_{\mr{eddy,1}}$ is the number of eddy rotations
necessary to explain the observed $\epsilon_{\msc{b}}$, the
requirement $N_{\rm eddy}t_{\rm eddy}\leq t_{\rm dec}$ for a wind-like
or ISM-like circumburst medium gives respectively
\begin{equation}
\label{eq:tconstraint} 
\left\{ \begin{array}{ll} L_{13}\frac{2}{\delta_{\max}}\lesssim0.3\,E_{\mr{iso, 53}}\,v_{\msc{wr},8.3} 
\,\dot{M}_{\msc{wr}, -5}^{-1}\,\Gamma^{-3}_{0,2}\,N_{\mr{eddy,1}}^{-1}
& \textrm{for $\bar{\rho}_0\propto r^{-2}$}\\ 
L_{13}\frac{2}{\delta_{\max}}\lesssim8.7\,(E_{\mr{iso, 53}}/n_{\msc{ism},0})^{1/3}\,\Gamma^{-5/3}_{0,2}
\,N_{\mr{eddy,1}}^{-1}
& \textrm{for $\bar\rho_0\approx\mbox{const.}$} 
\end{array} \right. 
\end{equation}
Let us emphasize that, for smaller initial shock Lorentz factors $\Gamma_0$, a larger size $L$ and weaker overdensity $\delta_{\mr{max}}$ are enough to satisfy the constraint $N_{\rm eddy}t_{\rm eddy}\leq t_{\rm dec}$.

\subsection{Vortical energy fraction}
The kinetic energy density invested in the vortical part of the post-shock flow should be a significant fraction of its proper energy density, since---as stated in \S II---we assume that the  
amplified magnetic fraction, which is required to match the observational inference $\epsilon_{\msc{b}}\gtrsim10^{-3}$, will eventually be comparable to the vortical fraction, $\epsilon_{\msc{b}}\sim\epsilon_{\rm{vort}}$. However, $\epsilon_{\msc{b}}$ might be smaller than $\epsilon_{\mr{vort}}$ if the backreaction of magnetic field on turbulent motions is important well before equipartition between the magnetic and vortical energy densities; on the other hand, $\epsilon_{\mr{vort}}$ may also be a lower limit for $\epsilon_{\msc{b}}$ if secondary shocks created by the overlapping sound waves of many different overpressured clumps significantly contribute to magnetic field amplification.

\begin{figure}[htbp]
  \centering 
\includegraphics[width=5in]{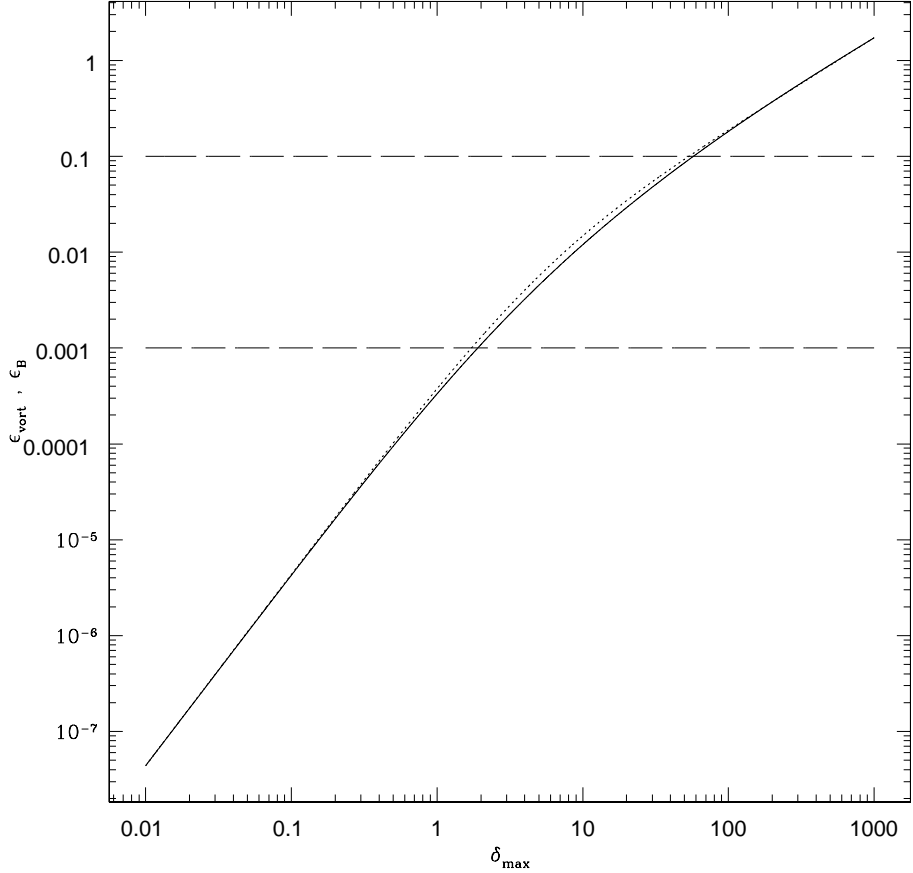}
\caption{
  \textit{Solid line}: Full numerical calculation of the vortical
  energy fraction $\epsilon_{\mr{vort}}$ as a function of the central
  overdensity $\delta_{\mr{max}}$, assuming an average shock Lorentz
  factor $\bar\Gamma=100$ and a clump volume filling factor $N_{c,
    0}L^3=0.25$ (a different choice for this parameter would simply
  shift the curve). \textit{Dotted line}: Analytical best-fit with the
  fitting function in eq.~({\ref{eq:fit}}) and fitting parameters
  $c_1\approx0.176$ and $c_2\approx1.054$. The area between the
  \textit{dashed lines} gives the region allowed by afterglow
  observations for the magnetic energy fraction $\epsilon_{\msc{b}}$.
}
  \label{fig:plot2} 
\end{figure}

The results of a numerical calculation for the vortical fraction of clumps with overdensity profile in eq.~(\ref{eq:density}) are shown in Figure~\ref{fig:plot2} (\emph{solid line}), but for small central overdensities the leading-order analytical computation described above gives a reliable estimate of the magnetic energy fraction produced by turbulence in GRB afterglows:
\begin{equation}\label{eq:energylimit}
\epsilon_{\msc{b}}\sim\epsilon_{\rm{vort}, \msc{lo}}\approx1.8\times10^{-3}\,\bar\Gamma_{2}^{-1}\left(\frac{\delta_{\rm{max}}}{2}\right)^2\frac{N_{\rm{c,0}}L^3}{0.25}
\end{equation}
where $\bar\Gamma=10^2\,\bar\Gamma_2$ and the reference values for the
central overdensity $\delta_{\mr{max}}$ and the clump volume filling
factor $N_{\rm{c},0}L^3$ have been chosen in order to match the
observational constraint $\epsilon_{\msc{b}}\gtrsim10^{-3}$. So,
clumps with moderate density contrasts ($\delta_{\rm max}\approx2$, in
agreement with the observational evidences reviewed in \S III) can
justify the lower limit of the magnetic energy fraction inferred from
afterglow models. Higher density contrasts (see
Figure~\ref{fig:plot2}) would be necessary to achieve larger magnetic
fractions; however, it is worth recalling that our model is applicable
only under the assumptions that the clump overdensity is not larger
than $\sim\Gamma$ (see \S II) and that $\delta_{\rm{max}}N_{\rm{c,0}}
L^3\lesssim1$, so that the overdensity profile can actually be
described by eq.~(\ref{eq:density}).

Eq.~(\ref{eq:energylimit}) suggests that smaller overdensities would
be enough to satisfy the observational constraints on
$\epsilon_{\msc{b}}$ as the forward shock slows down, since
$\epsilon_{\msc{b}}\propto\bar\Gamma^{-1}$; the magnetic fraction is
then expected to change during the afterglow stage, and in \S V we
will discuss the possible observational consequences of its time
evolution.

\section{Summary and Discussion}\label{sec:sum}

We have proposed that the post-shock magnetic fields
of GRB afterglows may arise from macroscopic MHD
turbulence rather than microscopic plasma
instabilities.  The source of turbulence is vorticity produced when
the shock encounters density inhomogeneities in the pre-shock medium.
We presume that the magnetic energy fraction
($\epsilon_{\msc{b}}$) that results is comparable to the energy fraction
of the turbulence.
The ultra-relativistic Geometrical Shock Dynamics
formalism of Paper I \citep{goodman_macfadyen07} allows an easy,
though approximate, calculation of the vorticity produced by a given
density inhomogeneity in the limit that the shock Lorentz factor
$\Gamma\gg 1$.

In this picture, the observational inference that
$\epsilon_{\msc{b}}\gtrsim 10^{-3}$ constrains both the amplitude and
the lengthscale of inhomogeneities.  Eqs.~(\ref{eq:fit}) and
(\ref{eq:energylimit}) roughly relate the total energy fraction in
vortical motions to the volume filling factor and density contrast of
the clumps; this energy fraction must be comparable to the inferred
post-shock $\epsilon_{\msc{b}}$.  Filling factors and density
contrasts of order unity are required when the shock is still highly
relativistic.  Eq.~(\ref{eq:tconstraint}), on the other hand, express
the constraint on the lengthscale and density contrast of individual
clumps (independently of their volume filling factor) so that the
eddies can wind the magnetic field up to the observed value in less
than the expansion time of the shock.  This second constraint favors
small lengthscales, so that the clumps responsible for field
amplification would probably be too small, at least individually, to
modulate the afterglow light curve.  Actually, \citet{nakar_granot06}
have recently shown that any pre-shock inhomogeneity can hardly be
responsible for the rebrightenings observed in some afterglow light
curves.

There is a question whether a fluid treatment of the post-shock flow is
justified at all since the plasma is collisionless.  The same question
arises in supernova remnants, to which the standard answer is that
magnetization of the plasma ensures a short effective mean free path.
The present case is more extreme because the particles are
relativistic, the lengthscales on which fluid-like behavior is required
are smaller, especially in the present work, and the pre-shock field
is energetically negligible.  The relativistic Larmor radius of the
post-shock ions based on the compressed ambient field is $r_{\msc{l}}\approx
\Gamma m_p c^2/eB\approx 10^{12}(3\unit{\,\mu G}/B_0)\unit{cm}$,
where $B_0$ is the pre-shock field strength.  This is smaller than the
maximum tolerable clump size for field amplification at the beginning
of the afterglow phase, though only barely so
(\S\ref{subsec:turnover}).  Furthermore, whether or not the Weibel
instability can produce persistent magnetic fields, it should
enforce fluid-like behavior by isotropizing particle distribution
functions whenever counterstreaming plasmas overlap.

A basic conclusion of this work is that vortical turbulence becomes
easier to produce with decreasing shock Lorentz factor.  Both the
energy and timescale constraints become easier to satisfy as $\Gamma$
decreases (however, our ultra-relativistic approximations break down
as $\Gamma\to 1$).  Therefore, if the post-shock magnetic energy
density is produced by macroscopic turbulence, it is likely that
$\epsilon_{\msc{b}}$ will evolve as the shock ages, complicating the
task of drawing physical inferences about the GRB environment from the
observational data.  The abundance of early X-ray light curves
provided by the \emph{Swift} satellite has already led to models that
are more complicated than the rather simple theoretical description of
the sparser \emph{BeppoSAX} results \citep{Galama_etal98}.  The light
curves are not single power laws in time, but show breaks and
sometimes ``flares'', suggesting a need for extended energy input from
the central source \citep[and references therein]{Zhang_etal06}.  But,
to date, most modelers have assumed constant $\epsilon_{\msc{b}}$ and
$\epsilon_e$ (the post-shock energy fraction in relativistic $e^\pm$)
within individual afterglows, although these parameters are often
allowed to vary from one afterglow to another.

The effect of an evolving $\epsilon_{\msc{b}}$ depends upon the relationship
between the observed frequency and certain critical frequencies in the
synchrotron spectrum. Particularly important is the cooling frequency,
the doppler-shifted synchrotron frequency of a post-shock electron or positron
that radiates much of its energy on a timescale comparable to the age
of the shock.  For a pre-shock medium with mass density profile
$\bar\rho_m(r)\equiv c^{-2}\bar\rho_0(r)
=Kr^{-\omega}$ averaged over clumps, 
and for an adiabatic relativistic shock
with constant isotropically equivalent energy $E_{\rm iso}$
(notwithstanding the above-cited inferences from \emph{Swift} data),
the cooling frequency evolves as
(up to dimensionless constants of order unity)
\begin{eqnarray}
  \label{eq:nucool}
  \nu_{\rm cool}&\approx& \frac{e\,m_e}{(1+z_{\textsc{grb}})\,\sigma_{\textsc{t}}^2}
\left[\epsilon_{\msc{b}}\,\bar\rho_m(r_{\rm s})\right]^{-3/2}r_{\rm s}^{-2}\nonumber\\
&\approx& \frac{e\,m_e}{(1+z_{\textsc{grb}})\,\sigma_{\textsc{t}}^2}\epsilon_{\msc{b}}^{-3/2}
K^{-4/(4-\omega)}\left[\frac{E_{\rm iso}}{c}~
(1+z_{\msc{grb}})^{-1}t\right]^{(3\omega-4)/2(4-\omega)}\,,
\end{eqnarray}
where $r_{\rm s}$ is the shock radius, $t$ the astronomical observer's
time and $z_{\msc{grb}}$ the GRB cosmological redshift.  Evidently,
the shock energy and Lorentz factor scale out of the cooling frequency
when the latter is expressed in terms of the shock radius.  If one
could be confident that the early afterglow evolves in a freely
expanding wind ($\omega=2$), which seems \emph{a priori} likely in
collapsar models, then the evolution of $\epsilon_{\msc{b}}$ could be
inferred by measuring that of $\nu_{\rm cool}$.

Present evidence suggests that the cooling frequency lies below the
X-ray regime in the early afterglow phase.  This conclusion rests on
the usual assumption that the synchrotron-emitting electrons are
injected with a power law distribution of energies,
$N(\gamma)d\gamma\propto \gamma^{-p}d\gamma$ for $\gamma>\gamma_{\rm
  min}\gg 1$, in which $p>2$ so that the total energy is dominated by
electron energies near $\gamma_{\min}m_e c^2$, whose characteristic
observed frequency is $\nu_{\rm min}$.  The observed specific flux is
often described as a power law in time and frequency, $F_\nu\propto
t^{-\alpha} \nu^{-\beta}$, despite various breaks and the
aforementioned flares.  If synchrotron emission dominates and
$\nu_{\rm cool}>\nu_{\rm min}$ (slow-cooling regime), the spectral
index is $\beta=(p-1)/2$ if $\nu_{\min}<\nu<\nu_{\rm cool}$ and
$\beta=p/2$ if $\nu>\nu_{\rm cool}$.  It is believed that the
acceleration index $p$ is not much larger than $2$, perhaps $p\approx
2.2-2.3$, which is then consistent with the observed X-ray indices
$\beta\approx-1$ observed by \emph{Swift} \citep[and references
therein]{Zhang_etal06} only if $h\nu_{\rm cool}<1\mbox{ keV}$.
Evaluating eq.~(\ref{eq:nucool}) at the deceleration radius
appropriate for a Wolf-Rayet wind (see eq.~(\ref{eq:rdec})), we find
that the afterglow phase begins with a cooling frequency that is
plausibly consistent with this constraint:
\begin{equation}
  \label{eq:nucdec}
  h\nu_{\rm cool,dec}\approx 0.2\,(1+z_{\msc{grb}})^{-1}
(10^4\epsilon_{\msc{b}})^{-3/2}\,v_{\msc{wr},8.3}^{5/2}
\,\dot M_{\msc{WR},-5}^{-5/2}\, E_{\rm iso,53}\,\Gamma_{0,2}^{-2}\,~\mbox{keV}.
\end{equation}
Unless the cooling frequency passes through the observed band, one
cannot learn much about the evolution of $\epsilon_{\msc{b}}$ from
observations at $\nu>\nu_{\rm cool}>\nu_{\min}$, because (up to
numerical factors of order unity)
\begin{equation}
  \label{eq:nuFnu}
\nu F_{\nu} \approx 
\frac{\epsilon_e\, E_{\rm iso}}{4\pi d_L^2\,(1+z_{\msc{grb}})^{-1}\,t}
\left(\frac{\gamma}{\gamma_{\min}}\right)^{2-p}
\end{equation}
in this regime, where $d_L$ is the luminosity distance;
$\epsilon_{\msc{b}}$ enters the above expression only \emph{via} the
correspondence
$\nu\approx(1+z_{\msc{grb}})^{-1}(\gamma\Gamma)^2(\epsilon_{\msc{b}}\bar\rho_m)^{1/2}
e/m_e$ between observed frequency $\nu$ and electron Lorentz factor
$\gamma$, and hence is raised to the small exponent
$(p-2)/4\lesssim0.1$.  On the other hand, relation~(\ref{eq:nuFnu}) indicates
that the flux above the cooling frequency provides an excellent measure of the energy in
the electron population \citep{FreedmanWaxman01}.


\acknowledgments We thank Ed Jenkins and Bruce Draine for discussions
of small-scale inhomogeneities in the ISM, and Eli Waxman for
discussions of recent GRB afterglow observations and their
implications.  This work was supported in part by the NSF Center for
Magnetic Self-Organization at Princeton.

\newpage
\bibliography{urv}

\newpage

\newpage

\newpage

\end{document}